\documentclass[prl,reprint]{revtex4-1}
\usepackage{amsmath}
\usepackage{amssymb}
\usepackage{graphicx}
\usepackage{color} 
\usepackage{comment}
\usepackage{xcolor,soul,framed,caption}
\colorlet{shadecolor}{yellow}

\usepackage{graphicx}
\usepackage{dcolumn}
\usepackage{bm}
\begin{document}

\title{When Models Interact with their Subjects:\\ The Dynamics of Model Aware Systems}

\author{Dervis Can Vural}
\affiliation{University of Illinois at Urbana-Champaign, 1110 W. Green St. Urbana, IL, USA}
\email{dvural@gmail.com}
\date{February 19, 2011}

\begin{abstract}
A scientific model need not be a passive and static descriptor of its subject. If the subject is affected by the model, the model must be updated to explain its affected subject. In this study, two models regarding the dynamics of model aware systems are presented. The first explores the behavior of ``prediction seeking'' (PSP) and ``prediction avoiding'' (PSP) populations under the influence of a model that describes them. The second explores the publishing behavior of a group of experimentalists coupled to a model by means of confirmation bias. It is found that model aware systems can exhibit convergent random or oscillatory behavior and display universal 1/f noise. A numerical simulation of the physical experimentalists is compared with actual publications of neutron life time and $\Lambda$ mass measurements and is in good quantitative agreement.
\end{abstract}
\maketitle
\section*{Introduction}
The notion that the act of observation can alter the observed system is familiar to any contemporary scientist. Less familiar is the notion that a model can alter the system it aims to describe. Such model-system couplings may be substantial for cognitive\cite{chapanis,fehr}, social\cite{gergens,beaman,foon,hong,hacking}, or economic\cite{callon,miller,mckenzie} systems in which the constituent agents have full or partial access to the model that represents them. It is conceivable for example, that an analysis of word frequencies in spoken or written language\cite{leech} can alter the behavior of those who produce these words; an analysis of stock market\cite{bakker} can alter the behavior of stock traders; and an analysis of fashion\cite{sproles} can alter the behavior of those who influence fashion. The ``physical sciences'' do not seem to be free of model-subject interaction either, since experimenters are influenced by theory through confirmation bias\cite{jeng} and theorists build new models based on these experiments.

The interaction between subjects and their models has long been recognized in separate disciplines, and referred as ``the looping effect''\cite{hacking} in philosophy, ``the enlightenment effect''\cite{gergens} in psychology, and ``performativity''\cite{callon} or ``virtualism''\cite{miller} in economics. The model aware behavior in the physical sciences has been recognized too, and rightfully renounced as ``confirmation bias'' '\cite{greenwald,jeng}. Unfortunately, the cited considerations are very qualitative, and discipline-specific. Presently there exists no quantitative and generally applicable theory of model awareness. Without such a theory, it is not possible to ask whether the description of a model aware system will converge to a fixed point, or perpetually change while manipulating its subject. Neither is it possible to ask how the behavior of a system changes with increasing model awareness, or what universal properties, if any, do model aware systems have.

The present study aims to answer these questions, at least partially, by introducing two quantitative models of model awareness, cast as much as possible in a discipline-independent language. The first describes a population of prediction seeking or prediction avoiding agents who update their behavior at every time step depending on the current model that describes them. The second describes the behavior of a population of experimental physical scientists, who decide whether to publish or not depending on the proximity of their data to the current ``model''. In both the physical and social case, populations provide feedback to the models too, since models must be updated to explain the current state of the population. The outcomes are then compared against two behavioral experiments\cite{chapanis,fehr} qualitatively, and particle physics data\cite{pdg} quantitatively.

Even though real-life models usually involve elaborate verbal descriptions, and/or sophisticated mathematical machinery, in this study we will consider much simpler ones such as propositions regarding majority behavior, or averages of multiple publications regarding a physical quantity. As simple as they may be, it seems appropriate to refer to them as ``models'' since they are falsifiable descriptions with predictive power.

\section{Models}
\subsection{Model Aware Social Populations}

There has been a number of agent-based herding/anti-herding approaches in the literature used to explain a wide variety of social and economical phenomena (the interested reader is referred to \cite{samanidou}). The present approach is reminiscent of a Polya urn process\cite{scalas}, in the sense that we will be modifying an ensemble according to its sampled outcomes.

We consider a population of $N\gg 1$, in which an individual can be in either one of the states $A$ or $B$. These states can represent any opinion, property and behavior. At any given time $t$ the state of the population can be characterized by the fraction of $0<\phi(t)=A_t/N<1$ of individuals in state $A$. After each time step, a scientist performs a measurement by choosing $n$ random individuals ($1\ll n\ll N$) out of $N$ and publishes whether the majority of $n$ is $A$ or $B$. In connection with our introductory remarks we take any statement regarding majority behavior as a model; one example may be ``investors avoid risk''.  

When the model is published, each individual among $N$ may become ``aware'' of the model with probability $q\ll1$, and subsequently update his or her state. The state of those unaware of the model is assumed to remain unchanged. 

Two types of populations will be considered separately, defined in terms of how the aware individuals update their state: A \emph{prediction seeking population} (PSP) is one in which the aware individuals align their state with the prediction of the published model. For example, if the publication reports that ``most people are $A$'', then the aware $B's$ become $A$ in the next step (the aware $A$'s remain unchanged). A \emph{prediction avoiding population} (PAP) is one in which individuals anti-align their state. For example, if the publication reports that ``most people are $A$'', then the aware $A$'s flip their state to $B$ (the aware $B$'s remain unchanged).

Once the aware subpopulation updates its state, a subsequent scientist will be sampling, measuring and modeling a population of different nature. We will consider the Markovian dynamics of $\phi(t)$ over many such iterations. Unlike a Polya process, our ensemble is fixed in size, the number of modified agents is nondeterministic, and the modification is done according to a majority rule.

Let us first focus on a PSP. Given the state $\phi(t)=A_t/N$, the probability of having $k$ people who are A's in a sample of $n$ is given by the hypergeometric distribution,
\begin{align}
P_A(k)=\frac{\dbinom{A_t}{k}\dbinom{N-A_t}{n-k}}{\dbinom{N}{n}}
\end{align}
 Without loss of generality, suppose $n$ is odd. The probability that the majority of the scientist's measurement sample is $A$ is
\begin{align}
P(M_A|A_t)=1-P(M_B|A_t)=\sum_{k>n/2}^{n}P_A(k) 
\end{align}
After the publication, the probability $P(r)$ that $r$ people will become aware of the result is
\begin{align}
 P(r)=\dbinom{N}{r}q^r(1-q)^{N-r}
\end{align}
Given that there exists $r$ aware individuals, and given the majority of the sample is $A$, the probability that $u>0$ of them ($u<r$) to change into $B$ is
\begin{align}
P(u|r,M_A)=\frac{\dbinom{A_t}{r-u}\dbinom{N-A_t}{u}}{\dbinom{N}{r}} 
\end{align}
Thus, the probability that the population changes from $A_t$ to $A_{t+1}=A_t+u$ is given by the transition matrix
\begin{align}
 \Gamma_{A,A+u}&=\sum_{r>u}^NP(u|r,M_A)P(M_A|A_t)P(r)\nonumber\\
&\approx P(u|qN,M_A)P(M_A|A_t)
\end{align}
Similarly, the probability that $u>0$ individuals change in the other direction is
\begin{align}
\Gamma_{A,A-u}&=\sum_{r>u}^NP(u|r,M_B)P(M_B|A_t)P(r)\nonumber\\
&\approx P(u|qN,M_B)P(M_B|A_t)
\end{align}
where this time,
\begin{align}
 P(u|r,M_B)=\frac{\dbinom{A_t}{u}\dbinom{N-A_t}{r-u}}{\dbinom{N}{r}}.
\end{align}
We immediately see that if $A_t=N$, $P(M_A|A_t)=1$, $P(u|r,M_A)=1$ and $\Gamma_{N,N}=1$. The same holds true if $A_t=0$, and $\Gamma_{0,0}=1$. Thus, $A=1$ and $A=N$ are the absorbing states of the corresponding Markov Chain.

\begin{figure}[!ht]
\begin{center}
\includegraphics[scale=0.8]{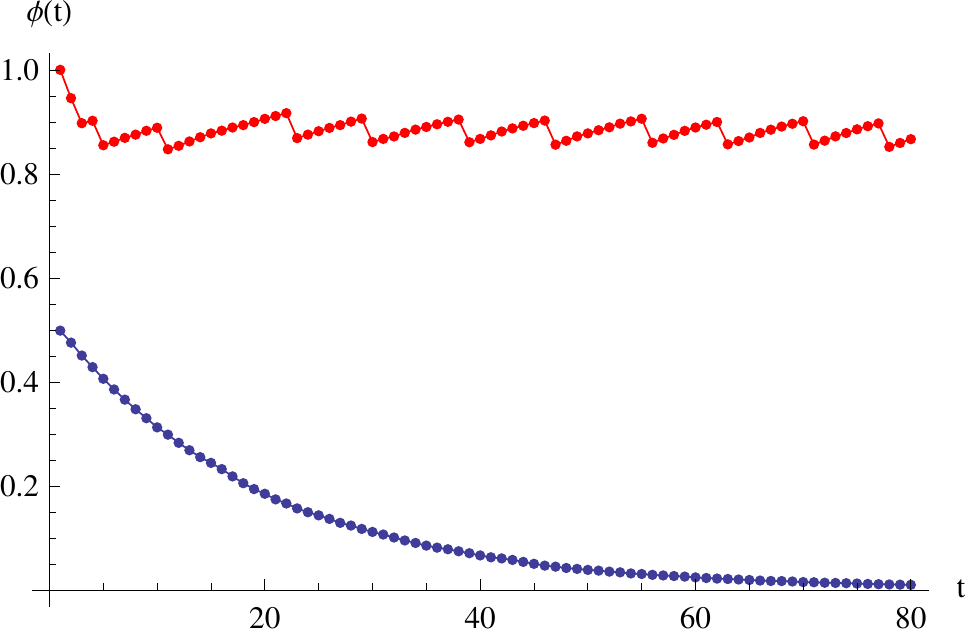}
\includegraphics[scale=0.8]{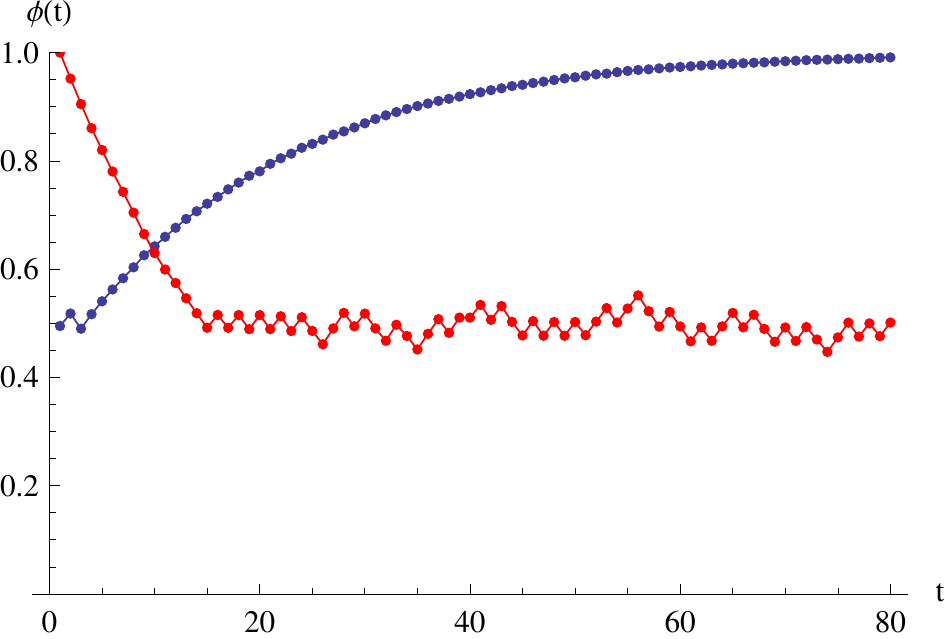}
\end{center}
\caption{
{\bf Fraction of individuals $\phi(t)$ having a certain quality $A$ as a function of time $t$.} Prediction seeking (blue) and prediction avoiding populations (red) evolve under the influence of an unbiased model (left) and a biassed e=0.45 model (right).  Here, $N=1\times10^4$, $n=5\times10^2$, $q=0.05$}
\label{Figure1}
\end{figure}

In the case of PAP, regardless of what the majority in the selected sample is, there always is a nonzero probability that the aware population will include some of individuals from the majority state, who in the next time step will flip. Any accidental trend due to finite sampling $n$ is counter balanced in the next time step by the prediction avoiders, who consequently cause an anti-trend. Thus, a PAP state will fluctuate on average by $q\phi$ (though $\langle\lim_{t\to\infty}\phi(t)\rangle$ exists; cf. below). The maximum expected change $\langle u_{max}\rangle=Nq$ will occur when $A_t=N$ or $A_t=0$. 

Let us now consider the expectation value $\langle \phi\rangle$. For a PSP the expectation value of the change $\langle \Delta\phi\rangle$ is $q(1-\phi(t))$ if $M_A$, and $-q\phi(t)$ if $M_B$. Therefore,
\begin{align}\label{1}
\langle\Delta\phi\rangle&=q(1-\phi)P(M_A|A_t)-q\phi(t)P(M_B|A_t)\nonumber\\&=q[P(M_A|A_t)-\phi].
\end{align}
Similarly, for a PAP $\langle\Delta\phi\rangle$ is $-q\phi(t)$ if $M_A$ and $q(1-\phi(t))$ if $M_B$. Therefore,
\begin{align}\label{2}
 \langle\Delta\phi\rangle=q[1-P(M_A|A_t)-\phi].
\end{align}
The function $P(M_A|\phi)<\phi$ for $\phi<1/2$, $P(M_A|\phi)=\phi$ for $\phi=1/2$ and $P(M_A|\phi)>\phi$ for $\phi>1/2$. Thus, the $\phi_{c}=1/2$ is a first order equilibrium state for both PSP and PAP. For a PSP, $\langle\Delta\phi\rangle$ is positive for $\phi>1/2$ and $\langle\lim_{t\to\infty}\phi(t)\rangle=1$ if $\phi(0)>1/2$. $\langle\Delta\phi\rangle$ is negative for $\phi<1/2$, and $\langle\lim_{t\to\infty}\phi(t)\rangle=0$ if $\phi(0)<1/2$. In contrast, for a PAP, $\langle\Delta\phi\rangle$ is negative for $\phi>1/2$ and positive for $\phi<1/2$; hence $\langle\lim_{t\to\infty}\phi(t)\rangle=1/2$ regardless of the initial state. Note that on average, the  variance $\Delta_t$ within the population is monotonically decreasing for a PSP, and monotonically increasing for a PAP (Fig(1)). 

It is also interesting to study the effects of incorrect models. This could happen for example due to an asymmetry in identifying one of the states, or misinterpreting correctly measured data during ``modeling''. Suppose that the scientist publishes $(1-e)\phi$ instead of $\phi$, the effect of which can be taken into account by modifying $P(M_A|A_t)$;
\begin{align}
P(M_A|A_t)'&=1-P(M_B|A_t)'\nonumber\\&=\sum_{k>n/[2(1-e)]}^{n}P_A(k)
\end{align}

By observing $\Delta\phi$ vs $\phi$ (cf. eqn(\ref{1}),(\ref{2}) and Fig(2)), we see that $e$ shifts the equilibrium point $\phi_c$ for both PAP and PSP. As a result, the PSP may now cross $\phi=0.5$, and unlike the $e=0$ case the variation within a PSP population need not monotonically decrease. The absorbing states of a PSP does not change, whereas for a PAP $e$ simply shifts the value of $\langle\lim_{t\to\infty}\phi\rangle$.

\begin{figure}[!ht]
\begin{center}
\includegraphics[scale=0.8]{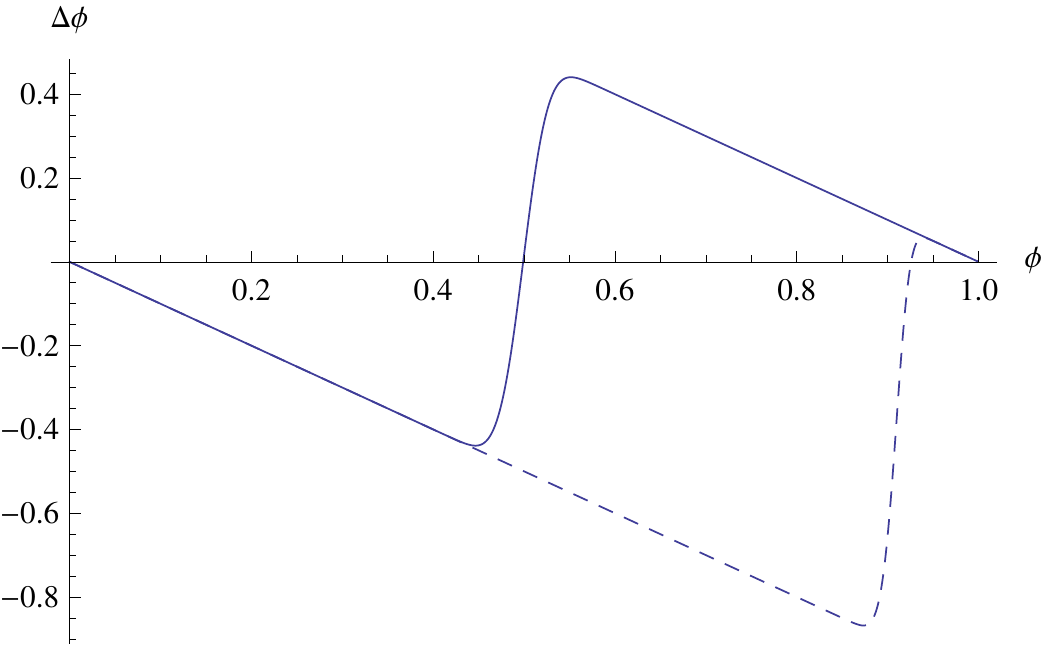}
\includegraphics[scale=0.8]{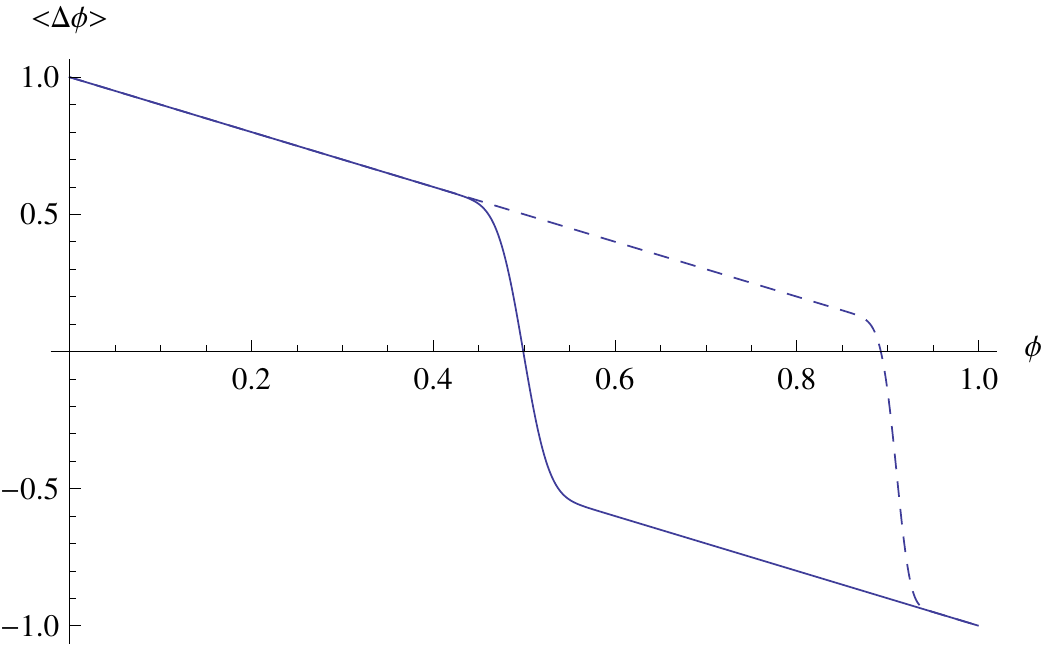}
\end{center}
\caption{
{\bf $<\Delta\phi(t)>$ as a function of $\phi$.} Prediction seeking (left) and prediction avoiding populations (right) in the presence of error $e=0$ (solid) and $e=0.45$ (dashed). $e$ shifts the equilibrium point for both PSP and PAP, and causes regular oscillations in PAP. Plot parameters are $N=1\times10^4$, $n=5\times10^2$, $q=0.05$.}
\label{Figure2}
\end{figure}

Numerical simulations are carried out for $N=10^4$, $n=500$, $q=0.05$, reveal interesting features regarding fluctuations (Fig(1), Fig(3)): First, in a PAP, as the error $e$ is gradually increased within $0<e<0.5$, the period of the small fluctuations in equilibrium dramatically increase and regularize (Fig(1)). The reason is evident in Fig(2); for $e=0$, we have $|\Delta\phi(\phi_c-\delta)|=|\Delta(\phi_c+\delta)|$, and therefore the period of trend/anti-trend fluctuations is equal to one time step. On the other hand for $0<e<0.5$, the difference in  $\Delta\phi(\phi_c-\delta)$ and $\Delta(\phi_c+\delta)$ leads to slower rises following rapid drops. As we continue to increase $0.5<e<1$, eqn(\ref{2}) now has a stable point, and the PAP starts to display convergent behavior much like the PSP. Thus, one could say that an unchanging model of a PAP is only possible if the model is practically wrong. 

A more interesting feature is that despite their opposite nature, PSP's and PAP's both appear to exhibit a $1/f$ noise in their power spectrum $P(f)$ if $q\ll1$ (Fig(3)). The $1/f$ noise is the signature of self-organized critical systems\cite{weisenfeld} and is ubiquitous in nature (see for eg.\cite{mandelbrot,voss,schick}). We predict that the same should appear in measurements of social model aware populations. 

\begin{figure}[!ht]
\begin{center}
\includegraphics[width=3.4in]{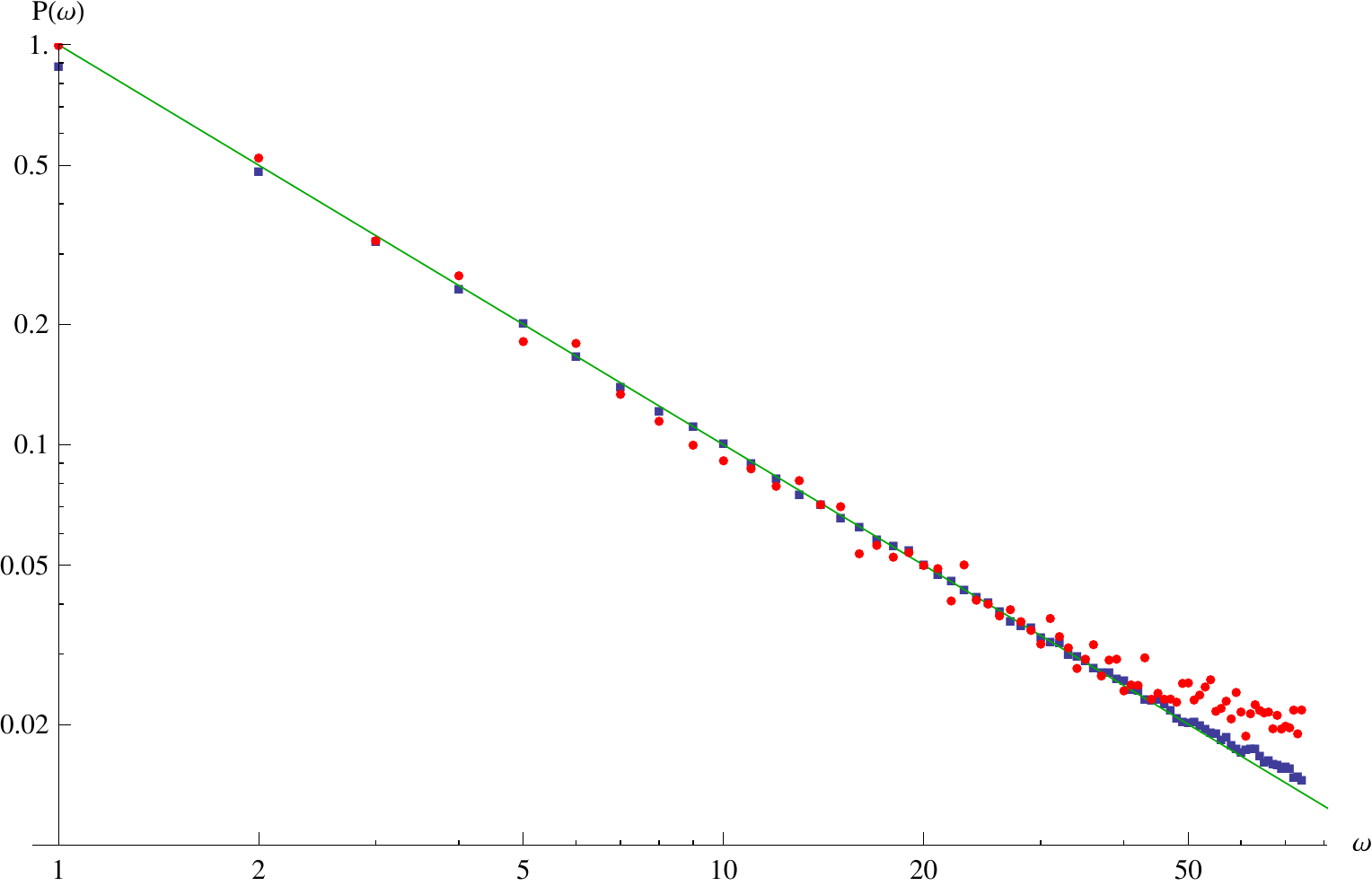}
\end{center}
\caption{
{\bf The power spectrum $P(\omega)$ of prediction avoiding (red) and prediction seeking (blue) populations}. It can be seen that both populations obey $1/\omega$ statistics (green). Here $N=10^4$, $n=500$ and $q=10^{-3}$.}
\label{Figure3}
\end{figure}
\subsection{The Model Aware Behavior of Physical Scientists}
Ideally, separate measurements of a physical quantity typically fluctuates around a fixed value according to a normal distribution due to random measurement errors. However, since experimenters tend to be influenced by past measurements and established theoretical models, data points are highly correlated\cite{jeng} and have visible systematic trends over time\cite{pdg}. This may be because, if the outcome of an experiment is significantly different from an accepted model or past outcomes, the experimenters may be ``improving'' their setup by eliminating some of the systematic errors which otherwise average out to zero, or even by simply repeating the experiment. In turn, future models are built on experimental data that have already been influenced by past models. Thus, it seems that the presence of model aware behavior extends beyond social sciences.

We describe the behavior of a population of model aware physical scientists with the following simplified characteristics: An experimenter measures a physical quantity $\mu$ at time $t$ and obtains $\phi(t)$ in order to publish the $n^{th}$ paper $P_n$ on the subject. Because of the precision limitations of the apparatus, $\phi(t)$ is normally distributed around $\mu$ with standard deviation $e_0$. After taking his measurement, the experimenter compares his result with the average $R_q(n)=q^{-1}\sum_{k=n-q+1}^nP_k$ of $q$ preceeding published measurements, with standard deviation $s_q(n)$.

In this case we define the ``model'' to be $R_q(n)$ (or any set of verbal or mathematical axioms that reproduce it). If $\phi(t)\pm e_0$ does not overlap with $R_q(n)\pm s_q(n)$, the experimentalist suspects there might be a mistake in the setup and decides to repeat the experiment at $t+1$. If on the other hand, if $\langle P\rangle_q\pm s_q$ overlaps with the ``model'', he publishes $P_{n+1}=\phi(t)$. It is assumed that $\mu$ is independent of time, and that non-random errors are negligible.

The random variable $P_{n+1}$ depends on the past $q$ publications $\{P_n, P_{n-1},\ldots,P_{n-q+1}\}$ and is distributed according to a ``partial Gaussian'' function.

\begin{align}\label{gaussian}
 \langle P_{n+1}\rangle=A(e,\mu,s,R)\int_{R_q(n)-s_q-e_0}^{R_q(n)+s_q+e_0}pf_{\mu,e_0}(p)dp
\end{align}
where $f_{\mu,e_0}(p)$ is the usual Gaussian distribution with mean $\mu$ and standard deviation $e_0$. Since the possible publications are limited by the scientist's confirmation bias, the normalization constant $A(e,\mu,s,R)$ is defined as $A\int_{R-s-e_0}^{R+s+e_0}f(p)dp=1$,

We can immediately see that when $s_q\gg e_0$, the lhs of (\ref{gaussian}) is precisely $\mu$ and $\langle P_{n+1}\rangle=\mu$ regardless of $P_n$; in other words, a lack of scientific consensus speeds up model convergence. The integral and $A$ can be calculated exactly,
\begin{align*}\label{scienceexpect}
\langle P_{n+1}\rangle=\mu+e_0\frac{e^{-(e_0+\mu-R+s)^2/(2 e_0^2)}-e^{-(e_0-\mu+R+s)^2/(2 e_0^2)}}{\sqrt{\pi/2} \left(\text{Erf}\left[\frac{e_0+\mu-R+s}{\sqrt{2} e_0}\right]+\text{Erf}\left[\frac{e_0-\mu+R+s}{\sqrt{2} e_0}\right]\right)}
 \end{align*}

Note that if $R(n)=\mu$ we get $\langle P_{n+1}\rangle=\mu$, and then $R(n+1)=R(n)$. Also, note that if $R(n)<\mu$ we have $\langle P_{n+1}\rangle>R(n)$, and as a result $R(n+1)>R(n)$. On the other hand if $R(n)>\mu$ we have $\langle P_{n+1}\rangle<R(n)$ and as a result $R(n+1)<R(n)$. Thus $R=\mu$ is a stable equilibrium point. Furthermore, the quantity $\langle P_{n+1}\rangle-R_n$ is linear in $-(R(n)-\mu)$ up to third order in $R(n)-\mu$, hence for values $R(n)\approx P_n\approx\mu$ we have exponential convergence (cf. Fig(4)).

\begin{figure*}[!ht]
\begin{center}
\includegraphics[scale=1]{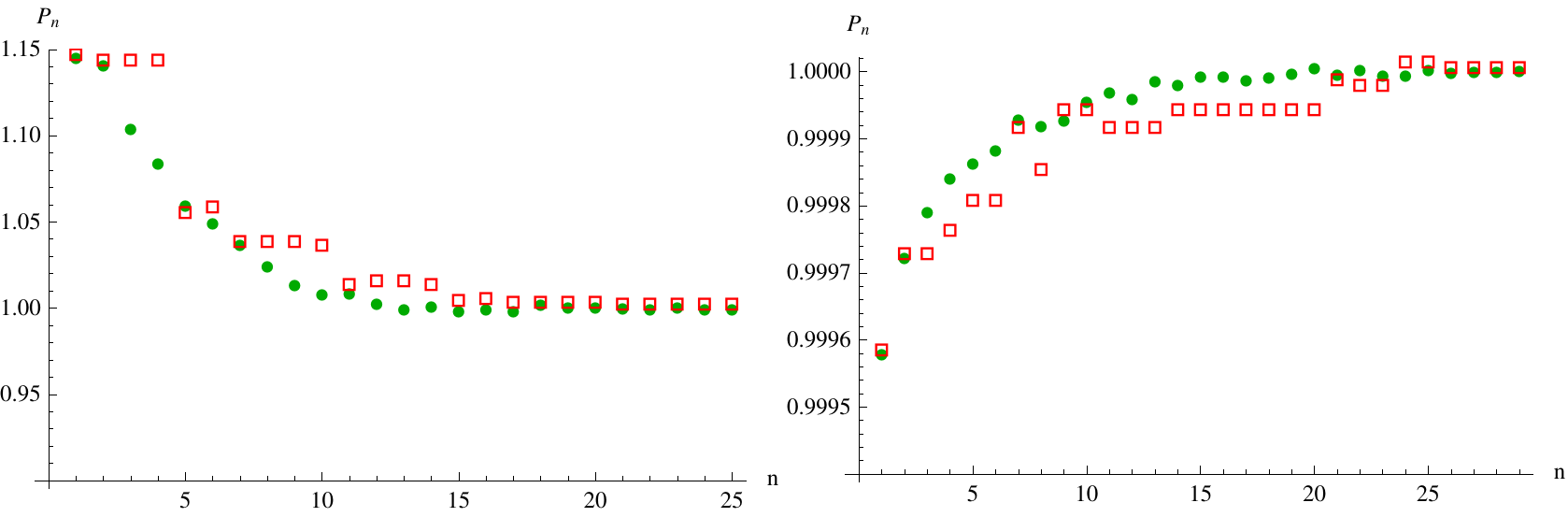}
\end{center}
\caption{
{\bf Published values of a physical quantity $P_n$ at publication number $n$.} Here, the Average values of publications of a simulated population of physical scientists (circles) is compared with actual \cite{pdg} particle physics measurements (boxes) of neutron life-time (left) and $\Lambda$ mass (right) normalized by their last value. $q=4$ is chosen for the neutron measurements and $q=8$ for the $\Lambda$ measurements. The two initial points are taken from data and used as initial conditions in the simulation.}
\label{Figure4}
\end{figure*}

Fig(4) shows the outcomes of exact numerical simulations and compares them with historical publications. The first two data points are taken from actual particle physics experiments \cite{pdg} as an input, and the rest of the data points are iterated. The short time behavior is characterized by tight clusters separated by abrupt jumps. The long time behavior approximately fits an exponential relaxation $P_n(\tau)=\mu+[\mu-e_0(0)]e^{-n/\tau}$ curve. The convergence time $\tau(q)$ determined from a least square fit, is a random variable with mean and standard deviation plotted as a function of $q$ (Fig(5)). The convergence times increase and diversify with increasing model awareness of the experimenters. We conclude that model aware experimenters considerably slow down scientific progress.
\begin{figure}[!ht]
\begin{center}
\includegraphics[scale=0.7]{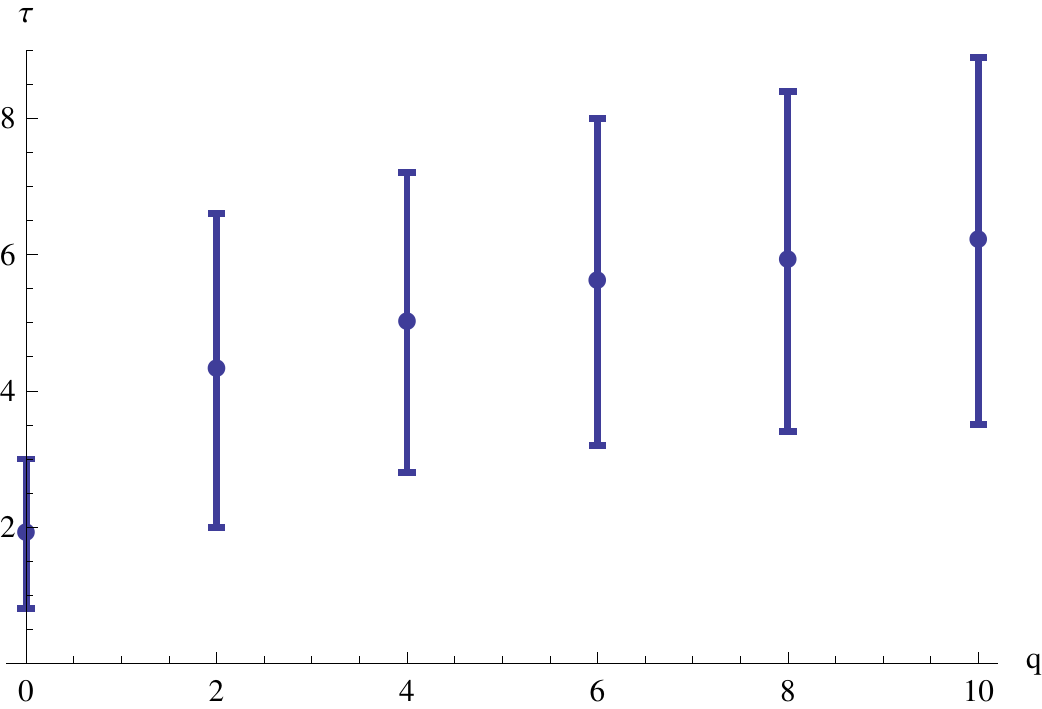}
\end{center}
\caption{
{\bf Average convergence time $\tau$ as a function of model awareness $q$.} Here two identical measurements with $15\%$ error is taken as initial conditions. The error bars indicate the standard deviation in $\tau$.}
\label{Figure5}
\end{figure}
\section{Discussion}
To further connect model awareness to the experimental literature, two recent behavioral studies \cite{chapanis, fehr} will be discussed within the PSP / PAP framework and will be compared with the simulation results. Due to lack of time dependent behavioral data, this comparison will have to be qualitative. Fortunately it will be possible to compare the outcomes of the simulations of the physical scientists quantitatively, since there exists records of historical particle physics publications regarding single quantities \cite{pdg}.

The first behavioral study is one on the behavioral effects of \emph{belief} of testosterone administration and actual testosterone administration \cite{fehr}. It was found that the women who \emph{believed} that they were given the hormone behaved significantly more unfair than those who \emph{believed} they were not given the hormone (regardless of whether or not they actually were). Curiously the effect of actual testosterone was to increase fairness. 

Measuring the effect of belief of testosterone administration relates to measuring the effect of the \emph{model} of testosterone on its model aware subject. The model of testosterone is the (as it turns out, incorrect) statement that ``testosterone causes people to be unfair''. The subjects of this study were familiar with this model, and behaved in accordance with its prediction. However, the study not only measures the effect of the testosterone model, it also alters it. Therefore if one repeats the same experiment in a few years, one may find that the effect of model to be the exact opposite, since if $q$ is large enough some participants may be familiar to the altered model of testosterone.

The second behavioral study is one on the effect of sophistication (defined as the knowledge of psychology and statistics) on the ability of generating random numbers \cite{chapanis}. Here it was found that sophistication can dramatically reduce -if not entirely eliminate- the many nonrandom trends unsophisticated people tend to display. Some such trends include avoiding repetitive triplets (such as 8,8,8) or favoring descending sequences (such as 5,4,3) over ascending ones (such as 3,4,5). Here too, the author measures the influence of a model, which in this case, describes how people generate random numbers. Some of the people, (labeled ``sophisticated'') are aware of this model, and actively avoid its predictions. However, now that some people are aware of how people generate random numbers, the model must be updated to included the behavior of this sophisticated fraction. Furthermore, if $q$ is large enough, upon repeating the experiment one might see the influence of ``super-sophisticated'' subjects, who avoid the trends that the mixed sophisticated-unsophisticated population display.

Unfortunately both of these experiments have only two data points that are useful for us. The belief of testosterone administration activates the present stage of a long model aware evolution in a PSP. Similarly the ``sophistication'' variable in the random number study is a marker of a similar final state of a PAP. Thus one may only compare $\phi_0$ to $\phi_t$ and $\Delta_0$ to $\Delta_t$, where $t$ is the present time. The short term behavior of the simulations qualitatively agree with both experiments \cite{chapanis,fehr}, the common features of which are (i) the model aware subjects are significantly different than the non model aware subjects and (ii) the non model aware subjects are more similar among themselves than model aware subjects.  

It is appropriate to represent the model aware subjects of the testosterone study (i.e. those who believe they received testosterone) by a PSP evolving under a strongly biased model. The model is taken to be biased because the statement that ``testosterone makes people unfair''  was demonstrated to be incorrect in the study. The simulation agrees with feature (i) and for short times, feature (ii). For long times, the simulated PSP eventually becomes less diverse than its starting state. Perhaps continued publications of studies similar to \cite{fehr} over the years, will reveal this time dependence. It is appropriate to represent the model aware subjects of the random number study \cite{chapanis} with a PAP evolving under an unbiased model. Here, too the simulation agrees with feature (i) and feature (ii) (remember that $\phi=0.5$ represents a maximally diverse population).

The actual neutron half-life and $\Lambda$-mass measurements from 1960 to 2010 as reported in \cite{pdg} are compared with the outcomes of the simulations, and is good agreement (Fig(4)). It is very interesting that other historical particle physics data reported in \cite{pdg} fits reasonably well to the same exponential form. The average convergence time $\tau$ depends on $q$. For example, since $\tau$ is larger for $\Lambda$ mass measurements compared to neutron life-time measurements, one can conclude that the scientists measuring $\Lambda$ mass were more model aware.

\section{Conclusion}
It was shown that for physical scientists and prediction seeking populations, models and their subjects can co-evolve to a consistent state. Loosely speaking, for these systems model awareness sets a lower time limit to reach ``truth''. For prediction avoiding populations, such a consistent state is possible only if the systematic error in the model is large enough ($e>0.5$) to make the model practically incorrect. For these systems, model awareness prevents convergence all together. This suggests that there may exist deterministic mechanical systems that are unpredictable due to their coupling to the predictor. The 1/f spectrum observed in populations of opposite nature suggest that there may be other universalities common to all model aware systems.

While this is just a preliminary study, the simulations presented here demonstrate that the models and their subjects can be highly coupled and radically alter each other. Since the interaction of subjects with models seem to be present in a very diverse range of fields, the framework proposed was intentionally kept simple. This way, the theorists of different fields can add relevant system-specific details, and study the variants of the proposed model. Such variants could include interaction between individuals, noise, or variable model errors.

Our study opens a wide range of additional questions that can be explored theoretically and experimentally: What is the influence of a time dependent model? Is it possible to construct a model that takes into account its own influence? For example, can a model predict its own acceptance in a community? Can the above considerations be generalized to more realistic models involving not mere numbers, but simulations or equations? How about model aware systems such as the stock market, where the subject is under the influence of multiple models? Hopefully our work will inspire the scientific community for a deeper exploration of model aware systems.

\section*{Acknowledgments}
I am grateful to P. Zorlutuna for numerous stimulating discussions, particularly for suggesting considering physical experimentalists; to F.T. Yarman Vural for suggestions and editing. I sincerely acknowledge one of the anonymous reviewer's insightful and constructive advices. I would like to thank A.J. Leggett for his support.

\appendix
\section{Appendix: Mathematica Codes}
\begin{verbatim}
(*The PSP/PAP Simulation*)

Clear[u, p, n, q, i, average];
u = 10000; (*total population*)
n = 500; (*experimenters sample population*)
steps = 150;(*total time steps*)
q = 0.01; (*awareness probability*)
p = Table[0, {i, 1, steps}];
average = Table[0, {i, 1, steps - 1}];
variance = Table[0, {i, 1, steps}];
ensemble = Table[0, {i, 1, steps}];
Flip[number_] := Mod[number + 1, 2];
Flipper[p_, average_, q_] :=
  
  Table[RandomChoice[{1 - q, q} -> {p[[i]],
      If[Abs[average - p[[i]]] <= 0.5,
         Flip[p[[i]]], p[[i]]]}], {i, 1, u}];
(*With probability q, flip deviants*)

(*For PSP use the below flipper instead*)

(*Flipper[p_, average_, q_] :=
  
    Table[RandomChoice[{1 - q, q} -> {p[[i]],
      If[Abs[average - p[[i]]] < 0.5,
         p[[i]], Flip[p[[i]]]]}], {i, 1, u}];
*)

(*Construct the initial population*)

p[[1]] = Table[RandomChoice[{0, 1} -> {0, 1}]
, {i, 1, u}];
(*measure the initial average
 via chosing n random people*)

variance[[1]] = Variance[RandomSample[p[[1]], n]];

(*iterate*)

For[i = 2, i <= steps, i++,
 picker = RandomSample[p[[i - 1]], n]; 
(*select n people*);
 average[[i - 1]] = Mean[picker]; 
(*Measure their average value*);
 variance[[i]] = StandardDeviation[picker]; 
(*Measure their variance*);
 p[[i]] = 
  Flipper[p[[i - 1]], average[[i - 1]], q]; 
(*Flip everyone accordingly*);
 ]
grandaverage = Table[Mean[p[[i]]], {i, 1, steps}];

(*Scientific Population Simulation*)
ensemblesize = 1000;
timerange = 26;
epsilon=10^{-7};
Clear[ensemble, w];
ensemble = Table[0, {i, 1, ensemblesize}];
For[w = 1, w <= ensemblesize, w++,
 Clear[i, j, p, data,
 publishprob, error, q, x, av, t, b];
 steps = 90;
 timerange = 26;
 trail = 4;
 p = Table[{0, 0}, {i, 1, steps}];
 average = Table[0, {i, 1, steps}];
 truth = 1;
 initial1 = 0.145;
 initial2 = 0.141;
 error[t_] := 
  If[t <= timerange, 
initial1 (1 - t/(timerange + 1)),epsilon];
 p[[1]] = {1, 1 + initial1};
 p[[2]] = {2, 1 + initial2};
 q = 1;
 j = 2;
 trailingdeviation = Table[0, {i, 1, steps}];
 trailingaverage = Table[0, {i, 1, steps}];
 trailingdeviation[[1]] = 
StandardDeviation[{p[[1, 2]],p[[2, 2]]}];
 trailingaverage[[1]] = (p[[1, 2]] + p[[2, 2]])/2;
 
 For[i = 2, i < steps, i++,
  data = RandomReal[NormalDistribution[truth, 
         error[j]]];
  trailingdeviation[[j]] = 
   If[j == 1 || trail == 1, 0, 
    If[j <= trail, 
      StandardDeviation[Table[p[[k, 2]], 
	      {k, 1, j}]], 
      StandardDeviation[Table[p[[k, 2]], 
         {k, j - trail + 1, j}]]]];
  trailingaverage[[j]] = 
   If[j <= trail, Mean[Table[p[[k, 2]],
      {k, 1, j}]], 
    Mean[Table[p[[k, 2]], 
      {k, j - trail + 1, j}]]];
   If[Abs[trailingaverage[[j]] - data] <= 
    q (error[j] + trailingdeviation[[j]]),
   j++;
   p[[j]] = {j, data}, True];
  ];
 p = Table[p[[i, 2]], {i, 1, j}];
 ensemble[[w]] = p
 ]
Show[ListPlot[Take[p, timerange - 1], 
  PlotRange -> {0.8, 1.4}, 
  Mesh -> Full, PlotStyle -> Red], 
  Plot[1 + (p[[1]] - 1) Exp[-(x - 1)/t], 
    {x, 1, steps}, 
    PlotRange -> All, PlotStyle -> Red]]
averages = 
 Table[Mean[Table[ensemble[[i, k]], 
{i, 1, ensemblesize}]], {k, 1, timerange}]
errors = Table[
  StandardDeviation[Table[ensemble[[i, k]],
      {i, 1, ensemblesize}]], {k, 1, timerange}]
ListPlot[averages, PlotStyle -> {Green}, 
 PlotMarkers -> {Automatic, 7}]
\end{verbatim}

\end{document}